\newcommand{\kms}{\rm ~km~s^{-1}}

\documentclass[12pt,preprint]{aastex}

\newcommand{\ha}{H$\alpha$}

\shorttitle{Radio and X-rays of SN~2001em}
\shortauthors{Chugai N.N. \& Chevalier R.A.}

\begin{document}
\title{LATE EMISSION FROM THE TYPE Ib/c SN 2001em:
OVERTAKING THE HYDROGEN ENVELOPE}

\author{Nikolai N. Chugai}
\affil{Institute of Astronomy, RAS, Pyatnitskaya 48, 109017 Moscow, Russia}

\and

\author{Roger A. Chevalier}
\affil{Department of Astronomy, University of Virginia, P.O. Box 3818,
Charlottesville, VA 22903, USA}
\email{rac5x@virginia.edu}



\begin{abstract}

The Type Ib/c supernova SN~2001em was observed to have
strong radio, X-ray, and H$\alpha$ emission
at an age of $\sim 2.5$ yr.
Although the radio and X-ray emission have been attributed
to an off-axis gamma-ray burst, we
model the emission as the
interaction of normal SN~Ib/c ejecta with
a dense, massive ($\sim3~M_{\odot}$) circumstellar
shell at a distance $\sim7\times10^{16}$ cm.
We investigate two models, in which
 the circumstellar shell has or has not been
overtaken by the forward shock at the time of the
X-ray observation.
The circumstellar shell was presumably formed by vigorous mass loss with a
rate $\sim(2-10)\times10^{-3}~M_{\odot}$ yr$^{-1}$ at
$\sim(1-2)\times10^3$ yr prior to the supernova explosion.
The   hydrogen envelope was completely lost, and subsequently
was swept up and accelerated by the fast wind of the presupernova
star up to a velocity of $30-50$ km s$^{-1}$.
Although interaction with the shell can explain most of the late emission
properties of SN 2001em, we need to invoke clumping of the gas to
explain the low absorption at X-ray and radio wavelengths.

\end{abstract}

\keywords{stars: mass-loss --- supernovae: general --- supernovae:
individual (\objectname{SN 2001em})}

\section{INTRODUCTION} \label{sec-intro}

The supernova SN 2001em was discovered on 2001 September 20 \citep{PL01}
in the galaxy UGC 11794 ($z=0.01935$).
With an apparent magnitude of about 18.6
and absolute magnitude of $M\approx-16$ at the time of discovery
(for $D=83$ Mpc), the supernova was present in an unfiltered image
on September 15 but not on September 5 at a level
of 19.5 mag \citep{PL01}.
The spectrum on 2001 October 20 was of Type Ib or Ic
a month after maximum brightness  \citep{FC01}.
There is little doubt
that SN~2001em was discovered early, probably before maximum light.
We adopt 2001 September 10, i.e. JD=2452163, as the time of explosion.
The 5 day uncertainty in the age is of no consequence for the
interpretation of the late observations considered here.

Two years after the explosion, on 2003 October 17,
SN~2001em was detected at radio wavelengths with the VLA at 3.6 cm.
Between then and 2004 January 30, the flux at 3.6 cm
increased by a factor of $\sim 1.3$  \citep{Sto04},
and by a factor of 1.56 to 2004 July 1  \citep{Sto05}.
The spectrum in the range $2-6.2$ cm was nonthermal,
$F_{\nu}\propto\nu^{-0.37}$ \citep{Sto04}.
The high radio luminosity $\sim2\times10^{28}$
erg s$^{-1}$ Hz$^{-1}$ at 6 cm  \citep{Sto04}
is unprecedented for a SN~Ib/c at this age.
Moreover, X-ray observations on 2004 April 4 (day 937)
with {\em Chandra} revealed X-ray emission in the
$0.5-8$ keV band with a luminosity of
$\sim 10^{41}$ erg s$^{-1}$  \citep{PL04}, again
unprecedented for a SN~Ib/c at this age.

The circumstellar (CS) medium in the immediate vicinity
of SNe~Ib/c is presumably
shaped by a Wolf-Rayet (WR) type wind. The
interaction with this wind is normally expected to
produce early radio emission that decays
with a fairly steep power law in time
\citep[e.g., SN~1983N,][]{Wei86}.
Although the emission would normally be undetectable at
an age of 2 years, \cite{Sto04} undertook
late observations of SNe Ib/c based on the suggestion
that strong late radio emission could be caused
by the interaction of a misaligned relativistic jet
with the WR wind  \citep[e.g.,][]{Pac01}.
\cite{GR04} discussed misaligned jet
and CS interaction models for the
late emission from SN~2001em and concluded that the
misaligned jet model was favored.
A prediction of their model was that the radio source should
be resolvable with VLBI (very long baseline interferometry)
observations, with an angular size $\ga 2$ mas.
VLBI observations have been undertaken by 2 groups
\citep{Sto05,BB05}, both of which failed
to resolve the source at the predicted size; Bietenholz \& Bartel
set an upper limit on the major axis angular size of 0.59 mas.
Although the VLBI observations do not support the misaligned jet
model, the model might not be ruled out if there is relativistic
motion of a compact radio source.

However, detection of a strong H$\alpha$ emission line
with FWHM of $1800\kms$
on 2004 May 7 \citep{SGK04} is not
readily explained in the misaligned jet model.
This type of emission is observed from supernovae which are
undergoing strong CS interaction and usually associated
with Type IIn supernovae  \citep[e.g.,][]{Fil97}.
The high radio and X-ray luminosities of SN 2001em
at an age $\sim10^3$ days are comparable
to those of bright Type IIn supernovae (e.g., SN 1986J, SN 1988Z)
at a similar age. Given this similarity, we
propose a model of SN 2001em that accounts for
the observed phenomena by interaction of a SN~Ib/c with a
dense CS shell that is some distance from the progenitor
and was initially part of the H envelope of the progenitor star.

General aspects of our model are discussed in \S~\ref{sec-general}
and more details are in \S~\ref{sec-model}.
The model results are compared to observations of SN~2001em
in \S~\ref{sec-result}
and the formation of the circumstellar shell is treated
in \S~\ref{sec-origin}.
The conclusions, with attention to the evolutionary status of
the SN~2001em progenitor, are in \S~\ref{sec-conc}.

\section{GENERAL CONSIDERATIONS}\label{sec-general}

We assume that SN~2001em exploded as an ordinary
SN~Ib/c, and the late X-ray, radio and \ha\ emission
were the outcome of the ejecta interaction with
a dense CS hydrogen shell lost by the progenitor star.
This conjecture implies that the CS environment around SN~2001em
was shaped by two episodes of mass loss:  heavy mass
loss in a slow red supergiant wind (possibly in a common envelope phase
or a superwind phase)   
and a subsequent rarefied, fast WR wind.
This WR wind caused the formation of a dense CS shell by the
interaction of the fast wind with a slow dense wind
\citep[e.g.,][]{Kah83}.

The WR stage following the loss of the hydrogen
envelope was relatively brief, so the stellar mass could not have
decreased significantly at this stage. We expect,
therefore, that the SN~2001em presupernova was the
He core of a massive star. A related example
is SN~1993J, which
originated from a ``typical'' mass range
$13-16~M_{\odot}$ and got rid of almost all the H envelope
\citep{Woo94}.
For SN~2001em, we adopt an ejecta mass of $M\approx 2.5~M_{\odot}$
and an energy of $E\approx1.6\times10^{51}$ erg, similar to
parameters found for SN~1993J  \citep[e.g.,][]{Utr94};
in \S~\ref{sec-dyn}, we consider variation of these parameters.
With the adopted parameters, the typical velocity of SN material is
$v_{\rm sn}=(2E/M)^{1/2}\approx8\times10^8$ cm s$^{-1}$.
Assuming that  X-rays from SN~2001em at $t\approx 937$ d
detected by {\em Chandra}  \citep{PL04} correspond to
the stage of  strong interaction with the CS shell, the
radius of the CS shell is then
$R_{\rm cs}\sim v_{\rm sn}t \sim 6\times10^{16}$ cm.  The typical
supernova velocity (or less) must be chosen if the supernova energy
has substantially thermalized; on the other hand, the rising radio
flux implies that the interaction is not highly evolved.
The hydrogen envelope was presumably lost  during the red supergiant
stage with a velocity $\sim 10^6$ cm s$^{-1}$.
The CS shell was likely accelerated by the WR
wind to a velocity $u_{cs}\sim 20\kms$, so the age may be 
$R_{\rm cs}/u_{cs}\sim 10^3$ yr
(\S \ref{sec-origin}).

The shock interaction of freely expanding SN ejecta ($v=r/t$)
with a CS shell proceeds through the formation of a double shock interface
layer with the forward shock accelerating
the CS gas and the reverse shock decelerating SN ejecta.
The swept up hot gas between the two shocks is
responsible for X-rays, while the accelerated electrons and
amplified magnetic field in the interaction zone bring about the
synchrotron radio emission \citep{Che82,CL89}.
The \ha\ emission line with a
full width at half maximum (FWHM) of $\sim1800$ km s$^{-1}$
detected on day 970  \citep{SGK04} is probably emitted by
the CS gas accelerated in the forward shock, since the
SN ejecta are devoid of hydrogen. The \ha\ profile has
no apparent extended wings beyond a velocity of $2000$ km s$^{-1}$
\citep{SGK04},
which implies that the bulk of the line-emitting gas in the forward
shock moves with velocities $\leq 2000$ km s$^{-1}$.

Two possibilities for the origin of the
high velocity hydrogen in the forward shock are:
(i) a shock wave with a velocity  $\sim2\times10^3$ km s$^{-1}$
that passed through the smooth dense CS shell was radiative, so a cool
dense shell formed between the shock wave and the contact surface;
(ii) the dense CS shell was clumpy, so the
clouds were first shocked by slow radiative shocks, and then
fragments of the shocked clouds were accelerated in the forward
shock up to $\sim2\times10^3$ km s$^{-1}$, similar to
the scenario proposed for the \ha\ emission
in SN~2002ic  \citep{CCL04}.

The temperature of the forward shock with a
velocity  of $\sim2\times10^3$ km s$^{-1}$ is about
$5$ keV, far below the 80 keV estimated from {\em Chandra} observations
on day 940  \citep{PL04}. The reverse shock can be
much hotter.
As a result of the sudden collision of the rarefied outer SN layers with the
dense CS shell, the swept up shell at the SN/CS interface
is strongly decelerated because of a high
CS/SN density contrast $\rho_{\rm cs}/\rho_{\rm sn}$. The swept up shell
velocity
\begin{equation}
v_{\rm s}\approx v_{\rm sn}(\rho_{\rm sn}/\rho_{\rm cs})^{1/2}\,,
\end{equation}
where $v_{\rm sn}$ is the velocity of SN ejecta at the reverse shock,
is low and, consequently, the reverse shock velocity
$v_{\rm sn}-v_{\rm s}$ is high enough to provide a
high temperature of the shocked SN ejecta. An additional
factor that favors a high temperature at the reverse shock
is an average molecular weight of ejecta that is larger than
for normal cosmic abundances.
For the ionized He composition of SN~Ib/c, $\bar{\mu}=\rho/nm_p =1.33$, so
the velocity of the reverse shock
must be only $\approx5500$ km s$^{-1}$ to yield a shock
temperature of 80 keV.   With the velocity of the swept up shell
$\approx 2000$ km s$^{-1}$, the boundary ejecta velocity should
be $\approx8000$ km s$^{-1}$ at the time of X-ray observation.
The actual picture may be more complicated, because
the X-ray emission is a combination of the
radiation from both shocks with different temperatures.

We now estimate the mass of the shocked SN ejecta
assuming that the reverse shock is the dominant component.
With the
standard bremsstrahlung cooling function
$\Lambda=1.6\times10^{-27}T^{1/2}Z^2$, where
$Z=2$ is the ion charge for pure He, we find that the observed X-rays
with $T=9\times10^8$~K (i.e. 80 keV) and
luminosity $L_x\approx 10^{41}$ erg s$^{-1}$ require
an emission measure ${\rm EM}\approx5\times10^{62}$ cm$^{-3}$.
To estimate the mass of the shocked ejecta,
the volume of the reverse postshock layer between the reverse shock and the contact
surface should be determined. Let SN ejecta with a
power low density distribution $\rho=\rho_0(v/v_0)^{-k}$ for $v>v_0$ and
$\rho=\rho_0$ at $v<v_0$ collide with a dense narrow CS shell.
The ratio of the radius of the reverse shock to the radius
of the contact surface $\xi=r_{\rm rs}/R_{\rm c}$ in that case is
\citep{CL89}
\begin{equation}
\xi=\left(\frac{4k-20}{4k-15}\right)^{1/3}\,.
\end{equation}
Substituting $k=9$ gives $\xi\approx0.91$.
For $R_{\rm c}\sim 6\times10^{16}$ cm,
$\xi=0.91$, and ${\rm EM}\approx5\times10^{62}$ cm$^{-3}$,
the mass of the shocked ejecta is $\sim 0.8~M_{\odot}$.
Momentum conservation implies that
the CS shell mass required to decelerate $M_1\sim 1~M_{\odot}$ of
the SN ejecta with the typical velocity of
$v_{\rm sn}\sim 8000$ km s$^{-1}$ down to $v_{\rm s}\sim 2000$ km s$^{-1}$
is $M_2\approx M_1(v_{\rm sn}/v_{\rm s}-1)\approx 3~M_{\odot}$.
Given the age of the CS shell of $\sim10^3$ yr, the last episode of the
hydrogen envelope removal thus occurred with a mass loss rate of
$\sim 3\times10^{-3}~M_{\odot}$ yr$^{-1}$.
Below we present a more detailed interaction model that will confirm
this general picture, although with some modifications regarding the 
interpretation of the X-ray spectrum (\S \ref{sec-dyn}).

This consideration of the X-ray luminosity shows that the mass of the
CS shell must be comparable to the ejecta mass, so that the supernova
energy is substantially thermalized.
Although the radio luminosity was rising at an age of 770--1000 days, 
we expect
that the luminosity cannot rise much further because the interaction
energy cannot keep increasing.
This is consistent with the fact that the luminosity of SN 2001em is
comparable to that of the most luminous radio and X-ray supernovae at
a comparable age.
The situation can be compared to that in SN 1987A, which is known to have
a dense ring at a radius of $6\times 10^{17}$ cm, about an order of magnitude
larger than the dense shell inferred here.
The radio flux from SN 1987A started to rise at an age of 3 yr
\citep{Man02}, but at
a luminosity level orders of magnitude smaller that the luminosity observed
in SN 2001em; the initial interaction is with the outer, high velocity,
low density ejecta, which have only a small part of the supernova energy.
In addition, the mass in the CS shell is probably considerably less than
the ejecta mass, so that only a fraction of the supernova energy will have
been thermalized at the time that the CS shell is swept up.

\section{THE MODEL}\label{sec-model}

\subsection{Hydrodynamic Interaction} \label{sec-interaction}

We consider freely expanding ($v=r/t$) SN ejecta interacting
with a CS environment consisting of a massive dense CS shell placed
between the WR wind (inner zone) and the slow red supergiant wind
(outer zone).
The SN ejecta mass is taken to be $M\approx 2.5~M_{\odot}$,
the kinetic energy is $E\approx1.6\times10^{51}$ erg, and the
density distribution has a flat inner zone, $\rho=\rho_0$ at $v<v_0$,
and a power law outer layer,
$\rho=\rho_0(v/v_0)^{-k}$ for $v>v_0$ with $k=9$.

\begin{figure}
\plotone{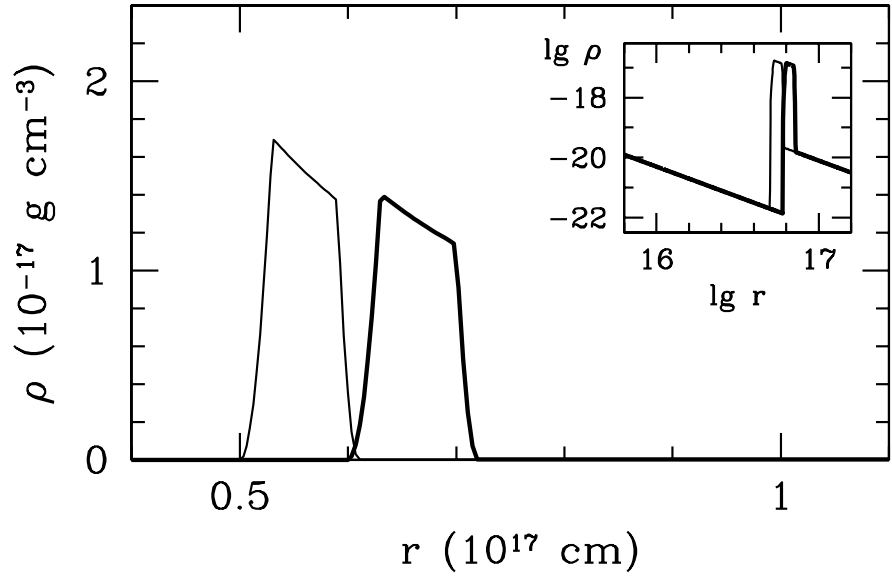}
\caption{The circumstellar density distribution for model A ({\em thick line})
   and model B ({\em thin line}) (see Table 1).
  Inset: the same distributions in logarithmic coordinates.
\label{f-density}}
\end{figure}

The WR wind is set by the wind density parameter
$w_1=\dot{M}_1/u_1=6.3\times10^{12}$ g cm$^{-1}$ which
corresponds to the choice $\dot{M}_1=10^{-5}~M_{\odot}$ yr$^{-1}$
and the wind velocity of $u_1=1000$ km s$^{-1}$.
This wind has little effect on the late time
interaction because of its low density, and we neglect
the fact that the
WR wind has to pass through a shock wave inside the CS shell.
The assumed width of the dense CS shell is
$R_2-R_1=\Delta R\approx 0.1R$,
where $R$ is the average radius of the CS shell.
The density in the CS shell is
characterized by the density parameter $w_2$,
which is defined by the shell radius and the shell mass with
the assumed $\Delta R/R$ ratio.
In \S~\ref{sec-den}, we show that the assumed shell thickness and density are
consistent with expectations.
The outer wind at $r>R_2$ is presumably the slow dense wind of
a red supergiant and we take $w_3=10^{15}$ g cm$^{-1}$,
an intermediate value between the wind densities expected for a SN~IIP and a
bright SN~IIL. We explored other values and found that, provided
$w_3\la 10^{16}$ g cm$^{-1}$, the results are not sensitive
to this parameter.
The two versions of the density distribution of the CS matter (CSM)
that will be used below are shown in Fig. \ref{f-density}.
The models differ in the radius of the CS shell.
In model A, the interaction with the
CS shell is  not yet finished
by the time of the X-ray observation ($t=937$ d), while
model B, with a smaller shell radius, represents a situation
where the CS shell was swept up prior to the X-ray observation.

The hydrodynamic evolution of the SN interaction
is treated in the thin shell approximation \citep{Che82},
i.e. the layer between forward and reverse shocks is replaced
by a thin layer. This description yields
the radius of the thin shell $R_{\rm s}$,
the shell velocity $v_{\rm s}$, and the boundary velocity of the
unshocked SN ejecta $v_{\rm sn}$ at an age $t$. The
velocities of the forward shock ($\approx v_{\rm s}$) and reverse
shock ($v_{\rm sn}-v_{\rm s}$) provide us with
kinetic luminosities for both shocks. The kinetic luminosities can
be converted into X-ray luminosities if they are multiplied
by a factor $t_{\rm e}/(t_{\rm e}+t_{\rm c})$ which is
determined by the ratio of the cooling time $t_{\rm c}$ and
the time of accumulation of the shock internal energy
$t_{\rm e}=(d{\rm ln}E/dt)^{-1}$.
Usually, $t_{\rm e}=t$ is assumed
\citep[e.g.,][]{CF94,Chu92} and this is appropriate
for interaction with a smooth wind. In the
case of a collision with a narrow CS shell one expects 
$t_{\rm e}<t$, so it is more appropriate
to use the directly calculated value of $t_{\rm e}$.
However, this approach based upon the instant kinetic
luminosity is not valid
at a very late epoch, when the forward shock has overrun the
dense shell ($R_{\rm s}>R_2$) and propagates into the rarefied wind.
The radiation from the large mass of
shocked CS gas left behind by the previous interaction
should dominate.
We treat the thermal history of the outer hot gas
by solving the time-dependent energy balance
with the internal energy generation in the
forward shock, adiabatic cooling (dominant term), and
radiative cooling. To obtain the density of the
outer hot gas, we assume that the shell expands homologously and 
$\Delta R/R=0.1$. This approximation is supported
by the fact that the spherical expansion of a gas cloud proceeds 
in a self-similar way with a boundary velocity comparable 
to the initial thermal velocity \citep{ZR67}, i.e. 
the shell speed in our case.  

The model simulations show that model A assuming a smooth CS density
predicts a larger absorption of X-rays than indicated by the
observations. Therefore, we assumed a clumpy structure of the CS shell
to reduce the absorption.
In model B the X-rays do not constrain the
structure of the CS shell,
because it has been already shocked at the time of observation.
We approximate the interaction with a clumpy CS shell
by including the clumpiness {\em only} for
the absorption computation, while
the interaction dynamics and the unabsorbed X-ray luminosity of both shocks
are described assuming a {\em smooth} CS shell with an average density.
This approximation, although crude at first glance,
is justified if CS clouds entering the forward shock fully
deposit their kinetic energy, which means that clouds are completely
crushed, fragmented and
mixed within the forward shock. This picture amounts to the
assumption that, for  CS clouds of radius $a$,
the cloud crushing time $t_{\rm cc}=a/v_{\rm c}$  \citep{KMC94}
is significantly less than the time required for a CS cloud to
cross the forward shock region with width $h$,
i.e., $t_{\rm cc}\ll (h/R)R/v_{\rm s}$. We return to this issue below
(\S \ref{sec-dyn}).

The question arises of  whether model A with a smooth density
can adequately represent the X-ray spectrum. The
spectrum of the forward shock in the cloudy CS gas
is a combination
of radiation from the hot intercloud gas and cooler cloud shocks,
which cannot be treated properly in the smooth approximation.
Fortunately, the X-ray emission of cloud shocks
does not affect markedly the amount of internal energy
generated by the forward shock in the intercloud gas. The point
is that the total
kinetic luminosity of slow cloud shocks is a factor of
$\chi_1^{-1/2}\ll1$ (where $\chi_1\gg1$ is cloud-to-intercloud
density ratio in the forward shock) smaller than the
kinetic luminosity of fast bow shocks.
Therefore, most of the internal energy in the forward shock region
resides in the hot intercloud gas.
The approximation of a smooth
forward shock region for the clumpy model A may be satisfactory
for the X-ray spectrum in the range $E>1$ keV.
Yet a sizable amount of the kinetic luminosity of
the forward shock interacting with a clumpy CS shell
is expected to be emitted by cloud shocks in the soft band ($E<0.5$ keV).
Most of this radiation is presumably absorbed and re-emitted in the
optical/UV band outside of the X-ray band of interest.

\subsection{X-ray and Radio Absorption Model}\label{sec-abs}

The observed X-ray spectrum is affected
by  four cool components, including (1) galactic absorption
and intrinsic absorption in (2) unshocked SN ejecta;
(3) unshocked CS gas; and (4) shocked cool \ha-emitting gas.
The reddening towards
SN~2001em is $E(B-V)=0.1$  \citep{SFB98}, which corresponds to
a column density
$N_{\rm H}=6\times10^{20}$ cm$^{-2}$ \citep{Spi78}. We take this
value as the total column density in the first cool component,
neglecting interstellar absorption in the host galaxy.
The SN ejecta can absorb X-rays, despite the fact that
ejecta gas can be heated up to $\sim6\times10^4$~K.
The smooth outer wind in the region $r>R_2$ irradiated by X-rays
becomes hot ($T>10^6$~K) for  $w_3\leq 10^{16}$ g cm$^{-1}$
and is not a significant absorber of radio. However the outer wind 
absorbs X-rays, although carbon and oxygen may be ionized up to 
He- and H-like ions. 

As noted above, the CS shell in model A
has to be clumpy to be consistent with the low X-ray absorption.
The optical depth of the clumpy CS shell is computed as
\begin{equation}
\tau(E)=\int_{R_{\rm s}}^{R_{\rm b}}
\pi a^2N_V\left(1-e^{-\tau_{\rm c}}\right)dr\,,
\label{eq-taux}
\end{equation}
where $R_{\rm b}$ is the outer radius of the CS region
taken to be $1.5\times10^{17}$ cm,
$N_V=f/(4\pi a^3/3)$ is the number density of clouds,
$f=(1-\xi)/\chi$ is the cloud filling factor,
$\xi$ is the mass fraction of the intercloud phase
(we take $\xi=0.1$) and $\chi$ is the ratio of cloud
density to the local average density,
$\tau_{\rm c}=(4/3)ak(E)\rho_{\rm c}$ is the cloud average optical
depth, and $k(E)$ is the absorption coefficient for X-rays with 
energy $E$.
The occultation optical depth of the CS
shell along the radius, $\tau_{\rm oc}=\pi a^2N_V\Delta R$, should
be of  order  unity or less to provide reasonable transparency. With
fixed $\xi$ and $\chi$, the only remaining parameter that
determines $\tau_{\rm oc}$ is the cloud radius $a$.
The X-ray absorption by the unshocked SN ejecta is determined by
integrating the radiation transfer equation along rays
with different impact parameters for the assumed density distribution.
We found that the absorption of radio emission by SN material has a
small effect on the emergent spectrum, so we treat it approximately 
in terms of the average values of the electron number density and the 
temperature of the ejecta irradiated by X-rays.

In model B, only the outer wind and fourth component (\ha-emitting gas)
can provide intrinsic X-ray absorption.
To treat the latter we assume that  $10-30$\%
of the shocked CS gas resides in the cool \ha-emitting gas;
the result is insensitive to the value in this range.
This gas
is probably distributed  in the form of heterogeneous structures
(knots, filaments, sheets) imbedded in the hot gas of the forward shock.
We assume that hot and cool gas are homogeneously mixed.
In this case the absorption by the fourth  component is characterized
by the optical depth $\tau_{4,\rm av}$ for the average column density,
and by the occultation optical depth (the average number of clouds along
the radius) $\tau_{4,\rm oc}$. The optical depth for the
X-ray radiation transmitted through the forward shock layer derived from
Eq. (\ref{eq-taux}) is then
\begin{equation}
\tau_4=\tau_{4,\rm oc}[1-\exp\,(-\tau_{4,\rm av}/\tau_{4,\rm oc})].
\end{equation}
For photons emitted in a layer along the normal to the surface,
the intensity of escaping radiation is $I=I_0[1-\exp\,(-\tau_4)]/\tau_4$,
where the $I_0$ is the unabsorbed intensity. We adopt the same
expression for the flux of radiation escaping from
layer $F=F_0[1-\exp\,(-\tau_4)]/\tau_4$.
However, for the radiation escaping through the inner surface
we additionally have to take into account the absorption in the
SN ejecta (e$^{-\tau_3}$)
and absorption in the \ha-emitting gas at the opposite side of the
forward shock (e$^{-\tau_4}$).
For the radio emission, we assume also that the \ha-emitting material is 
homogeneously mixed with the radio-emitting shell.
This is an approximation to a complicated situation 
in which the forward and reverse shock regions both contribute 
to the acceleration of relativistic electrons and magnetic
field amplification, so, generally, the radio-emitting shell and 
the \ha-emitting material may not overlap exactly. 

The free-free absorption of the radio emission by the CS shell
depends on the
ionization fraction of hydrogen ($x$) and the electron temperature
of the cool components. These
values are calculated taking into account that the absorbed energy
of X-rays is shared between Coulomb heating,
excitation and ionization. We assume that the latter two processes
have equal branching ratios. For the Coulomb heating we adopt
a branching ratio $x^{0.24}$, a reasonable approximation
if the ionization fraction $x>10^{-3}$  \citep{KF92}.
The cooling term in the energy balance is calculated using a
standard cooling function for  solar composition
\citep{SD93}. If the thermal balance cannot
maintain the cool phase ($T\leq10^5$ K) we adopt full ionization
and consider this gas as transparent.
The typical value of the ionization fraction of the clouds
at about 900 day is $x\approx0.3$ and temperature
$T_{\rm e}\approx13000-14000$~K,
while the intercloud gas is hot
($T_{\rm e}>10^6$~K) and thus does not absorb radio emission.
For the H$\alpha$-emitting gas the typical ionization is
$x\sim0.01$ and temperature $T_{\rm e}\approx 10^4$ K.
In the SN ejecta the helium is fully ionized and the
electron temperature is $T_{\rm e}\approx 6\times10^4$ K.

The unabsorbed  X-ray spectrum is modelled by thermal bremsstrahlung
with a non-relativistic Gaunt factor \citep{Ito00}.
The photoionization cross sections for the absorption
by electrons of K and L shells with $2s$ and $2l$ subshells
are taken from \cite{Ver93}. Solar
abundances are assumed for the CSM, while the SN~Ib/c ejecta are
approximated by a mixture of 65\% He, 33\% O, and 2\% Fe by mass .

\section{RESULTS}\label{sec-result}

\subsection{CS Interaction Models and X-ray Emission}\label{sec-dyn}

\begin{figure}
\plotone{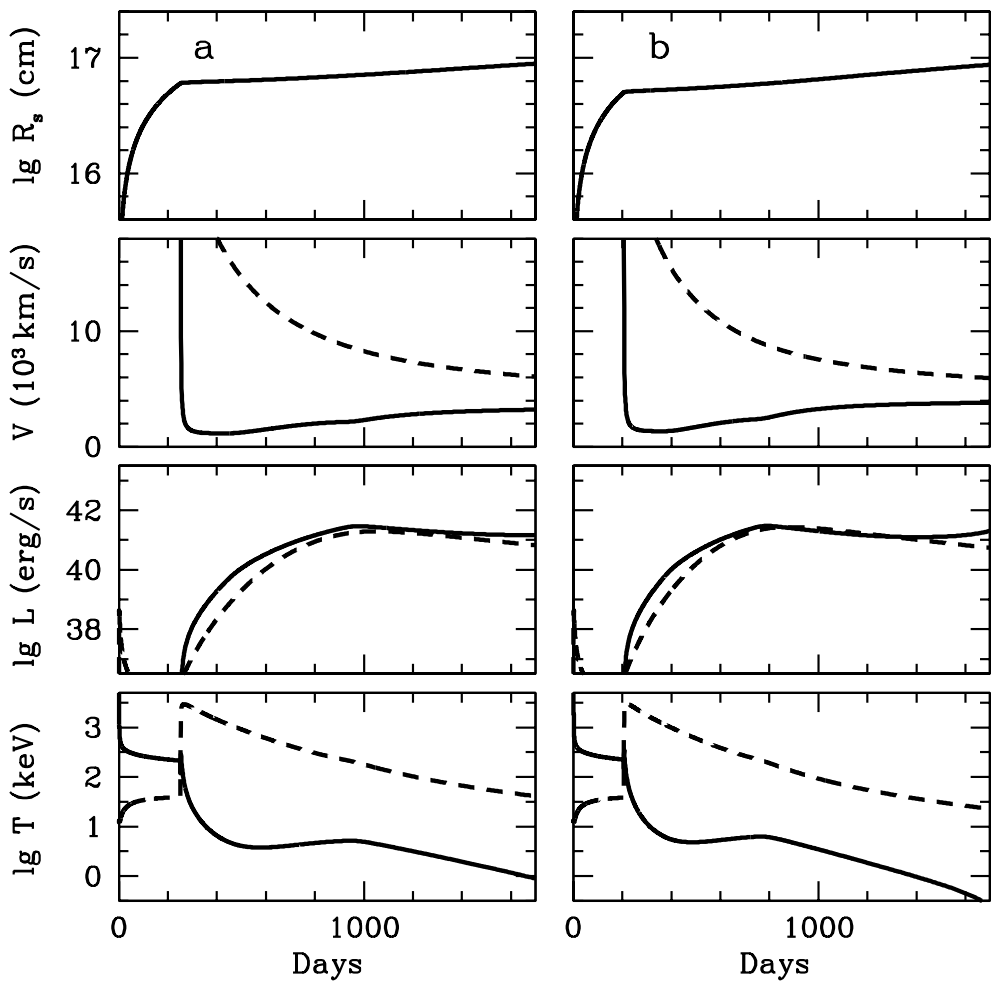}
  \caption{Evolution of the circumstellar interaction of SN~2001em:
  model A on the left (a) and model B on the right (b).
The plot shows from top to bottom the radius of the thin shell,
the expansion velocity of the
   thin shell ({\em solid}) and the SN ejecta velocity
   at the reverse shock ({\em dashed}),
  the unabsorbed X-ray luminosity of the forward shock ({\em solid})
  and reverse shock ({\em dashed}), and the temperature of the
  shocked CS gas ({\em solid}) and shocked SN gas ({\em dashed}).
  \label{f-dynamics}}
  \end{figure}

We now discuss models that fit the basic observations of SN 2001em.
The parameters of the CS shell for models A and B, namely, the
inner and outer radius of the CS shell ($R_1$ and $R_2$),
the density parameter of the CS shell ($w_2$),
and the total mass of the CS shell
are given in Table 1 (see also Fig. 1). The total mass of the CS shell
is $\approx 2.3-3~M_{\odot}$
(Table 1), similar to the preliminary estimate in \S~\ref{sec-general}.
The evolution of the major output parameters
is presented in Fig. \ref{f-dynamics}. The plot shows
the radius of the thin shell (contact surface), the velocity
of the thin shell and of the boundary of the SN ejecta,
the unabsorbed X-ray luminosity, and the temperature of forward and
reverse shocks. Here we assume a smooth CS shell so
both shocks are adiabatic; however, the forward shock 
between days 400 and 500 is very close to the radiative 
regime with $t_{\rm c}/t_{\rm e}\sim 1.5-2$.

After about day 200 a shell
formed during the interaction of the SN with the WR wind collides with the
inner boundary of the dense CS shell.
There is subsequent rapid
deceleration
from  $\sim 25000$ km s$^{-1}$ down to $1200-1300$ km s$^{-1}$,
followed by a period of steady acceleration \citep[e.g.,][]{CL89,Dwa05}.
The collision results in a rise of the X-ray luminosities
of both shocks toward their maximum values,  $\sim10^{41}$ erg s$^{-1}$.
The maxima  occur at the phase when the CS shell has been swept up,
i.e. at $\approx1000$ d and $\approx900$ d for models A and B
respectively.
The luminosity of the shocked CS gas remains high
after the CS shell has been overtaken; the temperature,  however,
rapidly decreases for $t>1000$ d due to adiabatic cooling.
The contribution of the forward shock to
the internal energy of the shocked CS gas at a late epoch
is negligible for $t<2000$ d.
At the time of the \ha\ observation ($t=970$ d), the thin shell velocity
is 2200 km s$^{-1}$ in model A and
3200 km s$^{-1}$ in model B; both are consistent with the observed \ha\
width.

\begin{figure}
\plotone{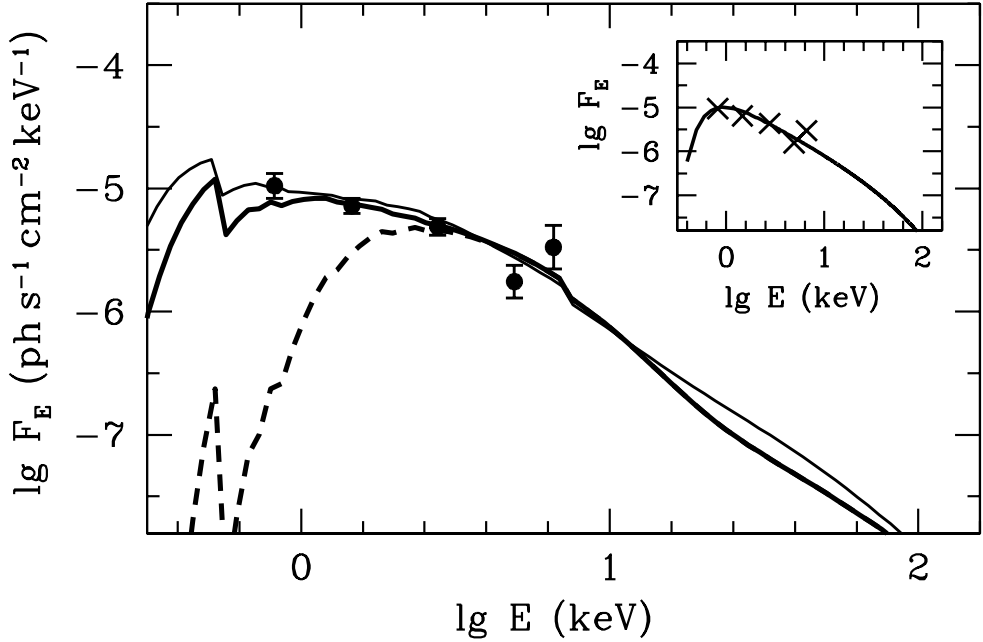}
  \caption{The photon spectrum of escaping X-ray radiation on day 937.
Both model A ({\em thick line}) with a clumpy CS shell
  and model B ({\em thin line})
  show satisfactory agreement with the observed fluxes (Pooley \& Lewin 2004).
  The models include the absorption by the clumpy
  cool \ha-emitting gas in the forward shock.
The fits are comparable to the fit of an
isothermal hot gas ($T=80$ keV) model with an external cool absorber
with $N_{\rm H}=1.6\times10^{21}$ cm$^{-2}$ (see inset, Pooley \& Lewin 2004).
Model A with a smooth CS shell ({\em dashed line}) shows unacceptably
  strong absorption.
   \label{f-xsp}}
  \end{figure}

The calculated X-ray spectra for both models (Fig. \ref{f-xsp})
reproduce the data of \cite{PL04} quite satisfactorily.
Note that model A has a clumpy CS shell here;
a smooth CS shell produces too strong absorption (Fig. \ref{f-xsp}).
The fit quality for both models is comparable to that of the
isothermal hot spectrum with $kT=80$ keV  (Fig. \ref{f-xsp}, inset) and with
external absorption corresponding to
$N_{\rm H}=1.6\times10^{21}$ cm$^{-2}$, the value
reported by \cite{PL04}.
The unabsorbed spectrum in our models is a mixture
of the hot ($kT\sim 100$ keV) radiation of the reverse shock and
``cool'' ($kT\sim5-6$ keV) radiation of the forward shock
with comparable luminosities (Fig. \ref{f-dynamics}).
In fact, the cooler component dominates in the range $\leq5$ keV.
We thus come to a picture in which the data are reproduced
by a relatively cool spectrum subjected to the intrinsic
absorption by the clumpy cool material, so the emergent
spectrum succesfully mimics the observed  hot isothermal spectrum
($\sim80$ keV) with a smooth external absorber.  Although
the general picture presented earlier (\S~ \ref{sec-general})
is based on the view that the X-rays are represented by
a hot isothermal spectrum, the mass estimate is not appreciably changed,
because the luminosity of the hot component is comparable to the total
luminosity.

The \ha-emitting gas
is characterized by an occultation optical depth
$\tau_{\rm oc,4}=0.5$ in model A
and $\tau_{\rm oc,4}=1.5$ in model B.
The cloudy CS shell in model A has $\tau_{\rm oc,2}=1$.
We adopt a cloud density contrast (cloud-to-average)
$\chi=10$, which leads to a cloud radius $a=4.7\times10^{14}$ cm.
The result is not sensitive to variation of $\chi$ in the range $3-30$.
With these parameters we can check the condition for the approximation of a
smooth density in model A, i.e.
$a\ll hv_{\rm c}/v_{\rm s}=a_0$ (\S~\ref{sec-interaction}).  From
momentum conservation, the ratio of the cloud shock velocity
to the forward shock velocity
$v_{\rm c}/v_{\rm s}\approx(4\xi/\chi)^{1/2}\approx 0.2$. Assuming
the forward shock width
$h\approx0.1R\approx7\times10^{15}$ cm, this gives
the upper limit $a_0=1.4\times10^{15}$ cm which is three times larger
than the cloud radius $a=4.7\times10^{14}$ cm in model A. The
condition for the smooth density approximation is thus barely met.

To consider variations in the supernova parameters, we
explored ``low energy,''
$M=2.5~M_{\odot}$, $E=10^{51}$ erg, and ``high mass,''
$M=4~M_{\odot}$, $E=1.6\times10^{51}$ erg, versions of model A.
We found that the
X-ray spectrum is well reproduced in both cases,
if the CS shell radius is smaller by a factor $0.8-0.84$.
The thin shell velocity is 1800 km s$^{-1}$ in the low energy case and
2000 km s$^{-1}$ in the high mass case, i.e. still consistent with
the \ha\ line profile. 
The uncertainty in the SN ejecta parameters thus results in 
minor variations in the model parameters.

\subsection{Radio Emission}\label{sec-radio}

The collision of the SN ejecta with the CS environment
is expected to produce synchrotron radio emission as a result of
the relativistic particle acceleration and magnetic field
amplification in the shock wave region \citep{Che82}.
As noted above, the situation with the radio evolution of SN~2001em is similar to
SN~1987A, in which  the radio flux showed steady growth after
the forward shock began to interact with the dense H II region
in the red supergiant wind \citep{CD95,Man02}.
Here we concentrate on the absorption effects of both free-free absorption
and synchrotron self-absorption,
and consider some implications of the radio evolution.
Our analysis is based on the assumptions that
relativistic electrons with a power law spectrum, $dN/dE=KE^{-p}$
(the spectral index is
$\alpha=(p-1)/2$ for $F_{\nu}\propto\nu^{-\alpha}$),
and magnetic field $B$
are distributed homogeneously in a spherical shell with  width
$\Delta R=0.1R_{\rm s}$ and radius $R_{\rm s}$; the radius is
provided by the interaction dynamics.
The radio-emitting shell is assumed to be homogeneously mixed with 
the cool clumpy \ha-emitting gas.
We assume a minimum energy of the relativistic electrons of 1~MeV,
energy equipartition between magnetic 
field and relativistic particles, and equipartition between relativistic
electrons and protons. Given the interaction model, the remaining
free parameter is the ratio ($\zeta$)
of the energy density of magnetic field plus relativistic particles to
the kinetic energy density of the forward shock.

\begin{figure}
\plotone{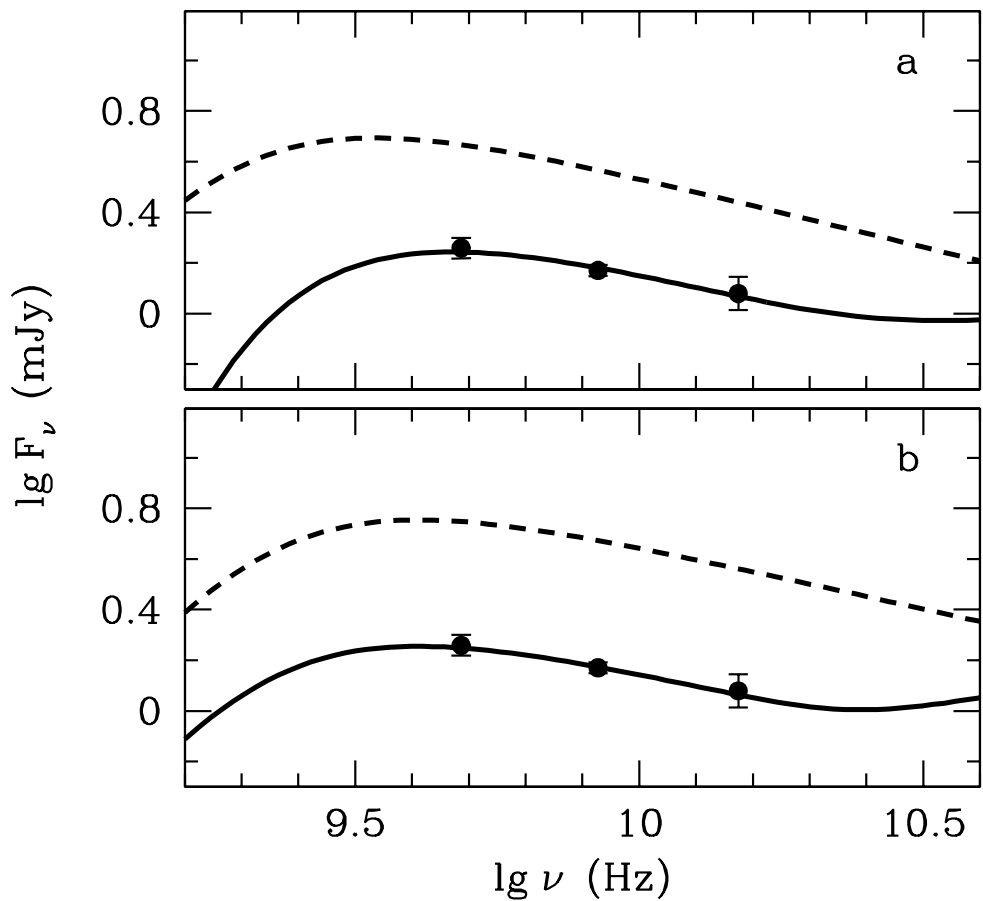}
  \caption{The model radio spectrum compared to
  the data of Stockdale et al. (2004) ({\em dots}). The {\em upper} panel
  shows the radio spectrum for  model A
  without   free-free absorption ({\em dashed line}) and with
  free-free absorption ({\em solid line}) produced by a
  clumpy CS shell and \ha-emitting gas in the forward shock.
  The {\em lower} panel is the same as the upper panel but
  for  model B and the free-free absorption produced only by
  \ha-emitting gas in the forward shock.
  \label{f-rsp}}
  \end{figure}

The effects of SSA (synchrotron self-absorption) and free-free absorption
on the radio spectrum in   model A and
 model B are shown in Fig. \ref{f-rsp} along with data
on 2004 January 31 \citep{Sto04}. The optimal parameters are $\alpha=0.55$,
$\zeta=0.024$ in model A and $\alpha=0.5$, $\zeta=0.017$ in model B.
We found that the variation of the SN ejecta parameters
in the low energy and high mass models (\S~\ref{sec-dyn})
results in a 10\% variation of the parameter $\zeta$ with
practically the same fit to the spectrum on day 872.
The derived values of $\zeta$ imply a
magnetic field $B\approx0.2$ G in both models at the time under
consideration. For this magnetic field, synchrotron losses
become important (i.e. $t_{\rm syn}\sim 1$ yr)
at frequencies $\nu>2\times10^{11}$ Hz. In the observed
frequency range ($\nu<2\times10^{10}$ Hz), 
the spectrum is not affected by synchrotron cooling.

The spectrum without free-free absorption demonstrates that
SSA is significant for $\nu<4\times10^9$ Hz.
The importance of SSA could be deduced from equation (13) of
\cite{Che98}, which shows that the size of the radio emitting
region at which the shell turns optically thin is comparable to the
size of the shell found here.
An observation of SN 2001em on 11 March 2005 at 1.6 GHz showed a considerably
lower flux than would have been expected from the higher frequency flux
evolution \citep{Par05}, which can be interpreted as a low
frequency turnover or a sudden flux decline.
Our results are consistent with a low frequency turnover.
The spectral inversion at $\nu>2\times10^{10}$ Hz is
related to the transparency of absorbing clumps at high frequencies.

The mechanisms for particle acceleration and magnetic field
amplification, especially in the case with a cloudy
structure of the CS shell, are uncertain.
The evolution of the radio flux, therefore, cannot be
predicted with confidence. Assuming that the energy of
relativistic electrons and magnetic field is a constant fraction
of the kinetic energy (i.e., $\zeta=const$), we find that
in model A the radio flux at 3.6 cm on day 767 is $\sim3$ times lower
compared to the observed flux, while on day 1025 the model flux
is $\sim2$ larger. This
implies that the parameter $\zeta$ should drop by factor of
three between days 767 and 1025. In model B assuming a smooth CS shell,
the flux on day 767 is 30 times lower owing to strong
absorption in the CS shell. On the other hand, assuming a clumpy
CS shell with the same clumpiness parameters as in model A,
we are able to reproduce in  model B the flux evolution between
day 767 and 1024 with $\zeta\approx const$.
This analysis indicates that model B with a clumpy CS shell
is somewhat preferred from the point of view of the
interpretation of  X-ray and radio data.

\subsection{\ha-emitting Gas}\label{sec-ha}

As remarked above (\S~\ref{sec-general}),
the \ha\ emission line on day 970 \citep{SGK04} indicates
the presence of cool shocked CS gas
behind the forward shock front.
We stress that in model B with the overtaken
CS shell, the forward shock is adiabatic for a smooth density
distribution and the assumed CS width $\Delta R/R=0.1$. The
shock becomes radiative if the density
is at least factor of three larger, i.e. $n\geq2\times10^7$ cm$^{-3}$.
This requires either a
factor of three narrower CS shell, i.e. $\Delta R/R\leq 0.03$,
or, alternatively, the CS shell could be clumpy. The latter
option is preferred because the
radio data, as noted above, also suggest a clumpy structure for
the CS shell in model B.
To be consistent with the \ha\ observation,
the cool gas in the forward shock of model B
must survive at least during $\sim100$ days after
it has been shocked. This is plausible,
although the survival of cool gas
in the hot environment is a complicated issue, because
it is related to the poorly defined mass exchange between
cool and hot phases, especially in the presence of magnetic field.

The dense shocked cool hydrogen re-emits the absorbed
X-rays,  producing \ha. The \ha\ luminosity
can be written as a fraction of the X-ray luminosity of the forward shock
absorbed by the fast cool hydrogen multiplied by the efficiency $\eta$
of the \ha\ emission.
The fraction of absorbed X-ray radiation with the bremsstrahlung spectrum
$F_E\propto E^{-0.4}\exp\,(-E/kT)$ \citep{Cox00}
is $\sim(E_1/kT)^{0.6}$, where
$E_1$ is the energy at which the optical depth of the cool hydrogen
shell with a mass $M_{\rm cool}$ is unity. An energy dependence
of the absorption coefficient $k_{\rm x}(E)\propto E^{-8/3}$ implies
$E_1\propto M_{\rm cool}^{3/8}$.
The expected H$\alpha$ luminosity
on day 970 with the temperature of X-ray radiation of the forward shock
$kT=5$ keV (Fig. \ref{f-dynamics}) is then
\begin{equation}
L_{{\rm H}\alpha}=5\times10^{40}\eta L_{{\rm x},41}\left(
\frac{M_{\rm cool}}{M_{\odot}}\right)^{0.225}\,,
\label{eq-lumha}
\end{equation}
where $L_{\rm x,41}$ is the X-ray luminosity of the forward shock in
units of $10^{41}$ erg s$^{-1}$.
The efficiency of \ha\ emission according to photoionization models
for relevant parameters,
i.e. the pressure equilibrium density of cool ($\sim10^4$~K)
gas of $\sim3\times10^{11}$ cm $^{-3}$,
column density of $\sim10^{22}$ cm $^{-3}$,
and X-ray flux of $\sim10^6$ erg s$^{-1}$ cm$^{-2}$ is
$\eta\sim 0.1$ \citep[][model 55]{CD89}.
The swept up CS mass in the forward shock on day 970 is
$2~M_{\odot}$ in  model A.
Assuming one-third of it resides in the cool phase
($0.7~M_{\odot}$)
the expected luminosity of \ha\ (Eq. [\ref{eq-lumha}])
is  then $\sim 4.5\times10^{39}$ erg s$^{-1}$ at about day 1000
for $L_{\rm x,41}=1$.
A comparable amount of a cool gas in model B would produce
a similar \ha\ luminosity. These estimates ignore the clumpiness 
of the \ha-emitting gas that might slightly reduce the luminosity. 
On the other hand, we also ignored  the effect of  thermal 
coductivity that might somewhat increase the \ha\ luminosity.
 Unfortunately, the observed \ha\ flux
is not yet available; its value would provide a useful
constraint on models.

An  implication of the SN/CSM interaction model
is the possible existence of a
narrow H$\alpha$ line from the ionized pre-shock CS gas.
For model A the luminosity of the narrow H$\alpha$ is
$\sim10^{38}$ erg s$^{-1}$ on day 1000, i.e.
$\sim 2$\% of the broad \ha\ line.    In
model B with the overtaken CS shell, a narrow
H$\alpha$ line is not expected. A search for narrow \ha\ emission
could provide a test for the presence of dense unshocked CS gas.

\section{FORMATION OF THE CS SHELL}\label{sec-origin}

\subsection{Age of the CS Shell}

The large mass and high density of the CS shell suggest
the following formation scenario:
loss of the hydrogen envelope  $\sim10^3$ yr
before the supernova,
possibly during a common envelope phase or the superwind
phase of a single star \citep{Heg97}, and the
subsequent sweeping up of this matter by
the fast wind of the  hot He star.
We thus envisage three major phases of the CS shell formation:
(I) mass-loss of the hydrogen envelope ($0<t<t_{\rm I}$);
(II) sweeping up the shell by the
WR wind ($t_{\rm I}<t<t_{\rm II}$); and (III) acceleration
of the swept up shell by the WR wind
($t_{\rm II}<t<t_{\rm III}$).
\cite{GLM96} have simulated this sequence of events, although for
different stellar parameters.
To estimate the age at the moment of the SN explosion
($t_{\rm III}$), we use the average parameters of our models:
final CS shell radius
$r_{\rm III}=6.5\times10^{16}$ cm and the mass of the CS
shell $M_{\rm cs}=2.7~M_{\odot}$.
The velocity of the mass-loss at stage I is assumed equal to
the typical red supergiant escape velocity $u_{\rm rsg}=10$ km s$^{-1}$,
while the mass-loss rate at this stage $\dot{M}_{\rm rsg}$
is a free parameter to be determined.
For the WR wind, the assumed parameters are
the same as stated above, i.e. $u_{\rm WR}=1000$ km s$^{-1}$ and
$\dot{M}_{\rm WR}=10^{-5}~M_{\odot}$ yr$^{-1}$.

The mass-loss rate at the first stage can be
constrained using the following arguments.
The minimum value of $\dot{M}_{\rm rsg}$ is determined
by the condition that at least the two first dynamical phases should
pass with the correct final radius of the CS shell
($r\approx 6.5\times10^{16}$ cm). The second phase is needed to
build the density of the CS shell up to a value
consistent with our models.
An upper limit of $\dot{M}_{\rm rsg}\approx 0.01~M_{\odot}$ yr$^{-1}$
is obtained assuming that the
average density parameter $w_2\approx 6.5\times10^{17}$ g cm$^{-1}$
in our models (Table 1)
corresponds to the unperturbed CS shell ejected by the presupernova.

The duration of the first phase is
$t_{\rm I}=M_{\rm cs}/\dot{M}_{\rm rsg}$. To estimate the time length
of phase II, we consider the interaction of two steady
winds. The interaction proceeds in the
pressure-dominated regime, i.e. the WR wind termination shock
has a relatively small radius.
In that case the swept-up shell expands with the
constant velocity $\lambda u_{\rm rsg}$ determined by the
cubic equation \citep{Kah83}
\begin{equation}
\lambda(1-\lambda)^2=\frac{1}{3}\frac{\dot{M}_{\rm WR}}{\dot{M}_{\rm rsg}}
\left(\frac{u_{\rm WR}}{u_{\rm rsg}}\right)^2.
\label{eq-kahn}
\end{equation}

In phase III, a thin shell of mass $M$ is driven
by the pressure of the shocked WR wind with the kinetic luminosity
$L_{\rm WR}$ and the internal energy of the shocked wind (bubble) $E$.
A similar problem for the
shell pushed by a pulsar wind has been solved earlier
\citep{RC84}.
The difference is in the adiabatic index: here we have
$\gamma=5/3$ instead of 4/3.
The equation of motion of the shell is then
\begin{equation}
M\frac{d^2r}{dt^2}=\frac{2E}{r}\,,
\label{eq-motion}
\end{equation}
while the energy equation is
\begin{equation}
\frac{dE}{dt}=L_{\rm WR}-\frac{2E}{r}\frac{dr}{dt}\,.
\label{eq-conserv}
\end{equation}
The solution of these equations is the self-similar
evolution of the CS shell radius
\begin{equation}
r=\left(\frac{2L_{\rm WR}}{3M}\right)^{1/2}t^{3/2}\,.
\label{eq-conserv-r}
\end{equation}
To apply this solution to the phase III of the CS shell expansion,
we identify the period $t_{\rm II}<t<t_{\rm III}$
with the time taken by the CS shell to expand in the
self-similar regime from the radius
$r_{\rm II}$ at $t=t_{\rm II}$ to the final
radius $r_{\rm \rm III}$.

For the parameters adopted in model B and the maximum mass-loss
rate during stage I
($\dot{M}_{\rm rsg}\approx 0.01~M_{\odot}$ yr$^{-1}$),
the phases I, II, and III end at $t_{\rm I}=260$ yr,
$t_{\rm II}=480$ yr, and
$t_{\rm III}\approx830$ yr respectively.
The CS shell velocity at the moment $t_{\rm III}$ is
$v_{\rm III}=55$ km s$^{-1}$.
The existence of the acceleration phase
$t_{\rm II}<t<t_{\rm III}$
provides in this case
a natural mechanism for the fragmentation of the CS shell
by the growth of the Rayleigh-Taylor instability \citep{GLM96}.
The minimum mass-loss rate at stage I,
determined from the requirement that
stage II should end just prior to the SN explosion,
is $\dot{M}_{\rm rsg}\approx 0.002~M_{\odot}$ yr$^{-1}$.
In this case $t_{\rm I}=1400$ yr, $t_{\rm  II}=2000$ yr, and
the shell velocity is 33 km s$^{-1}$.

We thus conclude that a major mass-loss episode with the rate
$(2-10)\times10^{-3}~M_{\odot}$ yr$^{-1}$ took place
1000--2000 yr before the SN explosion. The WR wind accelerated
this shell up to velocity of 30--50 km s$^{-1}$.

\subsection{Density of the CS Shell}\label{sec-den}

The minimum density in the CS shell predicted by our models
is $n\geq2\times10^7$ cm$^{-3}$ (\S~\ref{sec-ha}).
This value should reflect the pressure and thermal
balance at the latest phase of the
interaction of the WR wind with the CS shell.
The dynamical pressure created by the WR wind is
\begin{equation}
p_{WR}\approx\frac{\dot{M}_{\rm WR}v_{\rm WR}}{4\pi R^2}\,.
\label{eq-pressure}
\end{equation}
For the parameters adopted above
($\dot{M}_{\rm WR}=10^{-5}~M_{\odot}$ yr$^{-1}$,
$v_{\rm WR}=1000$ km s$^{-1}$, $R=6.5\times10^{16}$ cm) this
expression leads to $p_{WR}\approx10^{-6}$ dyn cm$^{-2}$.
Because $p_{WR}$ is actually determined by the pressure at the
WR wind termination shock, the pressure could be a factor
$\sim30$ higher than this estimate if the wind bubble is
contained within the shell.
However, if the bubble can break out of the shell because
of instabilities, the pressure is reduced.
We regard this pressure estimate as a lower limit.
To derive the density from pressure equilibrium
one needs to know the  temperature of the compressed CS gas.
First, we  check that the CS shell was mostly neutral.

The ionizing radiation of the presupernova forms a low density HII zone.
The density of the HII zone in pressure
equilibrium is $n_2=p_{WR}/kT_2\approx 7\times10^5$ cm$^{-3}$
assuming $T_2=10^4$ K. The column density of the
HII zone is
\begin{equation}
N_2=\frac{L_{\rm UV}}{4\pi R^2\alpha n_2{\rm Ry}}\,,
\label{eq-column}
\end{equation}
where $L_{\rm UV}$ is the luminosity of the ionizing radiation,
$\alpha$ is the hydrogen recombination coefficient, and Ry is
the hydrogen ionization potential.
The luminosity of the progenitor, assuming
a helium core of a $15~M_{\odot}$ star, is
$L=2\times10^{38}$ erg s$^{-1}$ \citep{Mey94}.
Adopting $L_{\rm UV}=0.5L\approx1\times10^{38}$ erg s$^{-1}$, $T_2=10^4$ K,
$\alpha=2\times10^{-13}$ cm$^3$ s$^{-1}$, and $R=6.5\times10^{16}$ cm,
we obtain $N_2\approx6\times10^{20}$ cm$^{-2}$.
This is substantially lower than the hydrogen column density of the
CS shell, $\sim6\times10^{22}$ cm$^{-2}$ for a CS shell
with a mass of $2.7~M_{\odot}$.
This estimate implies that the presupernova radiation ionizes
only a small fraction of the CS shell and
the bulk of the CS shell is neutral and can be referred to as
a photodissociation region \citep{TH85}.

The cool neutral CS shell with a column
density  of $N_{\rm H}\approx6\times10^{22}$ cm$^{-2}$ is characterized
by a dust extinction $A_V\sim10$, assuming interstellar
dust properties \citep{Spi78}.
In
a photodissociation zone with dust extinction $A_V\geq4$
the gas temperature follows the dust temperature owing to the
gas heating by the dust IR radiation \citep{TH85}.
With the emission efficiency of the dust at the maximum
of the blackbody infrared emission
$Q_{\rm e}=a/\lambda\approx3.4aT_{\rm d}$ (here $a$ is the grain radius),
the temperature of the dust grains should be
$T_{\rm d}\approx(L/55\pi R^2a\sigma)^{1/5}$.
With $L=2\times10^{38}$ erg s$^{-1}$, $R=6.5\times10^{16}$ cm
and  $a=10^{-6}$ cm, we obtain $T_{\rm d}\approx 300$ K.
Gas with this temperature in equilibrium with the pressure $p_{WR}$
has a density $n\sim2\times10^7$ cm$^{-3}$. This is
consistent with the lower limit for the density of CS shell clouds.
The suggested scenario of the CS shell formation
thus seems realistic.

\section{DISCUSSION AND CONCLUSIONS}\label{sec-conc}

Our goal was to propose a model for
the strong late time X-ray, radio, and \ha\ emission
from the Type Ib/c supernova SN~2001em.
We developed a picture in which the SN ejecta of a
normal SN~Ib/c collide with a dense massive CS shell
($M_{\rm cs}\sim3~M_{\odot}$)
at a radius of $\sim 0.7\times10^{17}$ cm.
We found that two scenarios are viable: model A in which the dense CS shell
not is yet overtaken by the time of X-ray observation (937 d)
and  model B with the  CS shell overtaken prior to the
X-ray observation.
The mass loss episode that led to the formation of 
the CS shell was characterized by a high mass loss  rate
$(2-10)\times10^{-3}~M_{\odot}$ yr$^{-1}$ at a stage
1000--2000 yr prior to the SN outburst. This material was
then swept up by the fast WR wind from the presupernova progenitor.
Rayleigh-Taylor instabilities likely led to the fragmentation of the
CS shell which is manifested in the low X-ray and radio
absorption despite the high column density of the shell.

The mechanism for the rapid loss of the
$3~M_{\odot}$ hydrogen envelope is not clear, although
it is probably related to  mechanisms for
producing SN~Ib/c presupernovae in a close binary scenario.
The bulk of SN~Ib/c presupernovae are
thought to originate by the stripping of the hydrogen envelope at
a late stage of binary star evolution as a result of
mass transfer to a companion and/or
the loss of the common envelope  \citep{van83,EW88,PJH92,WL99}.
In the case of SN~2001em, we believe that common envelope
evolution is the
most likely mechanism of the formation of the massive CS shell.
The $1000-2000$ yr delay between a strong mass loss episode
and the supernova explosion indicates that the former might
happen at the stage of carbon burning
in the core of a star with an initial mass in the range $14-17~M_{\odot}$
\citep{HMM04}. We suggest that the remains of the
hydrogen envelope ($\sim 3~M_{\odot}$) were lost
during the carbon burning phase due to the formation of
a common envelope in a binary system.

The relative fraction of SNe~Ib/c passing through the common envelope phase
is uncertain and lies between 10\%
\citep{PJH92} and nearly 100\%  \citep{TY02}. 
The massive star acquires a red supergiant structure at the
He burning stage.
Therefore, most SN~Ib/c presupernovae
pass through the common envelope stage long before ($\sim10^6$ yr)
the supernova explodes in
the rarefied CS environment formed by the WR wind. Only a small fraction
of SN~Ib/c presupernovae lose their H envelope at the C burning stage.
It was argued that this fraction is possibly about $\sim10^{-2}$ \citep{Chu97}.
We thus expect that roughly $0.001-0.01$ of all SNe~Ib/c
possess close  ($R\la 10^{17}$ cm) massive CS shells and
display CS interaction by
strong radio, optical and X-ray emission at an age of $1-10$ yr.
Upper limits on late radio emission from SNe~Ib/c
\citep{SNK05} seem to be
consistent with the estimated fraction of these events.

The occurrence of the supernova close to
the time that the H envelope was lost may have relevance
to some Type IIn supernovae with strong circumstellar interaction.
\citep{Fra02} found that the ejecta in SN 1995N inside
the reverse shock wave were heavy element rich,  implying that
this Type IIn supernova had lost essentially all the
hydrogen envelope in a dense wind at the time of the supernova.
It is unlikely, however, that the rate of all the Type IIn
supernovae can be explained entirely by the common envelope events
at the carbon burning phase \citep{Chu97}.

In addition to radio detection, an efficient way to find
SNe~Ib/c with close CS shells could be the detection
of late time \ha\ emission with a luminosity of
$\sim 10^{39}-10^{40}$ erg s$^{-1}$  at an age of  $1-10$ yr,
as has occurred for some Type IIn supernovae.
Another interesting way to detect the presence of the
massive close CS shell around SN~Ib/c is related to the
possibility that the shell is dusty. If the radius of the CS shell
exceeds the dust evaporation radius $\sim (1-3)\times10^{17}$ cm,
the observations of the infrared echo in the $KLM$ bands
may reveal the presence of a CS shell
soon after the SN~Ib/c explosion,
similar to infrared dust  echos observed
from SN~1979C and SN~1980K \citep{BE80,Dwe83}.
These observations will also serve to distinguish late circumstellar
interaction from misaligned gamma ray burst events, which must be
present in the population of SNe Ib/c at some level.

\acknowledgments
We are grateful to Dave Pooley for kindly sending us the
observed X-ray fluxes of SN~2001em.
This research was supported in part by Chandra grant TM4-5003X
and NSF grant AST-0307366.

{}
\clearpage

\begin{table}
\begin{center}
\caption{Model parameters}
\bigskip
\begin{tabular}{cccc}
\tableline\tableline
Parameter & Units  & A & B\\
\tableline
$R_1$            &  $10^{16}$ cm           & 6.3   & 5.3 \\
$R_2$            &  $10^{16}$ cm           & 7     & 5.9\\
$w_2$            &  $10^{17}$ g cm$^{-1}$  & 7     &  6\\
$M_{\rm cs}$     &  $M_{\odot}$            & 3.1   & 2.3 \\
\tableline
\end{tabular}
\end{center}
\end{table}

\clearpage

\clearpage

\clearpage

\clearpage

\end{document}